\documentclass[twocolumn,showpacs,aps,prl,superscriptaddress,letterpaper]{revtex4}
\usepackage{graphicx}
\usepackage{dcolumn}
\usepackage{amsmath}
\usepackage{epsfig}
\long\def\inst#1{\par\nobreak\kern 4pt\nobreak
    {\itshape #1}\par\vskip 10pt plus 3pt minus 3pt}
\RequirePackage{xspace}
\usepackage{relsize}

\def\pep2{PEP-II}

\def\vetoEff{$\epsilon_{v}$}
\def\splitoffEff{$\epsilon_{s}$}

\def\ee{$e^+e^-$}

\def\MeVc2 {${\rm MeV}/c^2$}
\def\GeVc2 {${\rm GeV}/c^2$}
\def\gevc {\rm GeV/c}
\def\gevc2 {${\rm GeV}/c^2$}

\def\babar{$\mbox{\sl B\hspace{-0.4em} {\small\sl A}\hspace{-0.37em}
\sl B\hspace{-0.4em} {\small\sl A\hspace{-0.02em}R}}$}

\begin{document}
  
  \begin{flushleft}
    \babar-PUB-07/028\\
    SLAC-PUB-12660\\
    [10mm]
  \end{flushleft}

  \title{
    \large \bfseries \boldmath Search for Prompt Production of $\chi_{c}$ and $X$(3872) in $e^+e^-$ Annihilations
  }

\date{\today}
%
\author{B.~Aubert}
\author{M.~Bona}
\author{D.~Boutigny}
\author{Y.~Karyotakis}
\author{J.~P.~Lees}
\author{V.~Poireau}
\author{X.~Prudent}
\author{V.~Tisserand}
\author{A.~Zghiche}
\affiliation{Laboratoire de Physique des Particules, IN2P3/CNRS et Universit\'e de Savoie, F-74941 Annecy-Le-Vieux, France }
\author{J.~Garra~Tico}
\author{E.~Grauges}
\affiliation{Universitat de Barcelona, Facultat de Fisica, Departament ECM, E-08028 Barcelona, Spain }
% ...
\collaboration{The \babar\ Collaboration}
\noaffiliation

\begin{abstract}
  We have searched for prompt production of $\chi_{c1}$, $\chi_{c2}$ and $X(3872)$
  in continuum $e^+e^-$ annihilations using a 386 fb$^{-1}$ data sample
  collected around $\sqrt{s} = 10.6$ GeV with the \babar\ detector
  using the $\gamma J/\psi$ decay mode.
  After accounting for the feed-down from $\psi(2S)\rightarrow\gamma\chi_{c1,2}$,
  no significant signal for prompt $\chi_{c1,2}$ production is observed. We
  present improved upper limits on the cross-section,     
  with the rest of the event consisting of more than two charged tracks, to be 
  77 fb for $\chi_{c1}$ and 79 fb for $\chi_{c2}$
  with $e^+e^-$ center-of-mass frame $\chi_c$ momentum greater than 2.0 GeV
  at 90\% confidence level.
  These limits are consistent with NRQCD predictions.  
  We also set an upper limit on the prompt production of $X(3872)$ through the decay
  $X(3872)\rightarrow \gamma J/\psi$. 
\end{abstract}


\maketitle

Charmonium production in $e^+e^-$ annihilation
provides opportunities to study both perturbative and
non-perturbative effects in QCD 
and to search for new charmonium states \cite{prompt_theory,dbltheory}.
The prompt production of $J/\psi$ and $\psi(2S)$ in $e^+e^-$
annihilation \cite{babarprompt,belleprompt}
and of double-charmonium \cite{belledbl,babardbl} have been observed at $B$-factory experiments.
These observations are surprising because the measured cross-sections are larger than
non-relativistic QCD (NRQCD) calculations by up to an order of magnitude
\cite{prompt_theory,dbltheory2}.

In the NRQCD production mechanism, a heavy quarkonium ($q \bar q$) state
can be produced at short distances as a conventional color-singlet,
or as a color-octet state, which then evolves into an observed quarkonium
meson along with other light hadrons. With this color-octet mechanism, one may explain the enhancement for $J/\psi$ production in \ee\ annihilation
\cite{prompt_theory}.
The production of $\chi_{c1,2}$ ($\chi_c$) in \ee\ annihilations
is an excellent probe of color-octet contributions, which are more prominent in
$\chi_c$ production than in $J/\psi$ production. This is because color-octet
and color-singlet processes enter $\chi_c$ production at the same order,
and $C$-parity suppresses the process $e^+e^-\rightarrow c\bar{c} gg$, which dominates
$J/\psi$ production.
Calculated cross-sections for prompt $\chi_c$ production in \ee\ annihilations are
$\sigma(e^+e^- \rightarrow \chi_{c1}  X) = 85$ fb and 
$\sigma(e^+e^- \rightarrow \chi_{c2}  X) = 123$ fb,
with \ee\ center-of-mass (CM) frame $J/\psi$ momentum
$p^\ast_{J/\psi} > 2.0$ GeV, and where
$X$ is one of $q\bar q$, $gg$ and $g$ in the leading order processes
and $J/\psi$ is from $\chi_c \rightarrow \gamma J/\psi$ decay
\cite{theory}.
More accurate measurements for $\chi_c$ states
will help to clarify the discrepancy between theoretical calculations and
existing measurements, and may point to other methods and mechanisms in
QCD to explain the differences.

Prompt production of charmonium mesons in $e^+e^-$ annihilation has been searched for
using either the reconstructed mass in an exclusive decay mode \cite{babarprompt,belleprompt}
or the mass distribution of
the system recoiling against the $J/\psi$ or $\psi(2S)$ \cite{belledbl,babardbl}.
Although prompt production of $\chi_{c0}$ has been observed, prompt 
production of the other $\chi_c$ states, $\chi_{c1}$ and $\chi_{c2}$, has not
been observed.

In this paper, we present a search for prompt $\chi_{c}$ production
in continuum $e^+e^-$ annihilation
using the $\gamma J/\psi$ ($J/\psi\rightarrow \ell^+\ell^-$) decay mode,
which is experimentally clean and is the dominant one in $\chi_c$ decay.
The current limits
on prompt production of $\chi_c$ are 
$\sigma(e^+e^-\rightarrow \chi_{c1} X) < 350$ fb
and
$\sigma(e^+e^-\rightarrow \chi_{c2} X) < 660$ fb
with \ee\ CM frame $\chi_c$ momentum $p^\ast_{\chi_c} > 2.0$ GeV,
where $X$ is the rest of the event \cite{belleprompt}. 
Belle and \babar\ recently observed an indication of the decay
$X(3872) \rightarrow \gamma J/\psi$ in $B$ decays \cite{x3872}, and therefore we
also search for prompt $X(3872)$ production using the
$\gamma J/\psi$ decay mode in $e^+e^-$ annihilation.

The data used in this analysis were collected with the \babar\ detector
at the PEP-II asymmetric-energy \ee\ collider, where
9.0 GeV electrons and 3.1 GeV positrons are collided at a
CM energy of 10.58 GeV, the mass of the $\Upsilon(4S)$
resonance.
The integrated luminosity ($\cal L$) 
consists of 349 fb$^{-1}$ (${\cal L}_{on}$) at the $\Upsilon(4S)$ resonance 
and 37 fb$^{-1}$ (${\cal L}_{off}$)
at a center-of-mass energy 40 MeV below the resonance.

The \babar\ detector is described elsewhere \cite{ref:babar} and 
here we give only a brief overview.
The momenta of charged particles are measured by the silicon
vertex tracker, consisting of five layers of double-sided silicon strip
sensors, and the central drift chamber (DCH)
with 40 wire layers, both operating in a
1.5 T magnetic field of a solenoid. The tracking system covers 92\% of the
solid angle in the CM frame. An internally-reflecting ring-imaging
Cherenkov detector (DIRC) with quartz bar radiators provides charged particle
identification (PID). A CsI(Tl) electromagnetic calorimeter (EMC)
is used to detect
and identify photons and electrons, while muons are identified in the
instrumented magnetic flux return system (IFR).

Electron candidates are identified by the ratio of the shower energy
measured in the EMC to the track momentum measured in the DCH,
the shower shape, the specific ionization energy loss in the DCH, and the Cherenkov angle measured by the DIRC.
Muons are identified by the depth of penetration into the IFR, the IFR
cluster geometry, and the energy deposited in the EMC. Photon candidates
are identified by EMC clusters that have a shape consistent with
an electromagnetic shower and are not associated with a charged track.

We use a Monte Carlo (MC) simulation of the \babar\ detector based on
GEANT4 \cite{GEANT4} to validate the analysis procedure,
to evaluate signal detection efficiencies, to model probability density
functions (PDFs), and to estimate background contributions.
We use samples of $e^+e^-\rightarrow \chi_{c1,2}\ +$ $J/\psi$
or $\psi(2S)$ MC events
to determine the selection criteria.
To estimate the signal reconstruction efficiencies and PDFs, we use
single $\chi_c$ MC samples decaying to $\gamma J/\psi$ with $J/\psi\rightarrow e^+e^-$
or $J/\psi\rightarrow \mu^+\mu^-$, which are generated with flat distributions in
$p^\ast$ (CM frame $\chi_c$ momentum) and
$\cos\theta^\ast$ (cosine of the polar angle of the $\chi_c$ momentum to the beam axis
in the CM frame).
To understand combinatorial background, we use MC generated 
$e^+e^-\rightarrow$ $\eta_c, \chi_{c0}$, or $\eta_c(2S)$ events produced in
association with either $J/\psi$ or $\psi(2S)$ mesons.
$B\overline B$ generic and initial state radiation (ISR) $\psi(2S)$
($e^+e^-\rightarrow \gamma \psi(2S)$) MC events
are used to estimate background contamination.
The $\chi_{c}$ candidates from $B$ decay
are used as a control data sample to correct for 
differences in the photon energy measurements 
between MC simulation and data. 

Charged particles are required to have
a point of closest approach to the beam spot of less than 10 cm 
along the beam axis and less than 1.5 cm in 
the plane transverse to the beam.
The $J/\psi$ mesons are reconstructed in the dilepton channel using two
oppositely-charged tracks identified as electrons or muons.
An algorithm to recover the energy loss due to bremsstrahlung
is applied to electron candidates.
The invariant mass of the reconstructed $J/\psi$ is required to
be within the range
[3.07, 3.13] GeV for the $\mu^+\mu^-$ channel and [3.05, 3.13] GeV for the
$e^+e^-$ channel. The asymmetric selection in the \ee\ channel is due to initial and final state radiation.
The $J/\psi$ candidate is
subjected to a vertex constrained fit and is combined with a
photon candidate that satisfies standard reconstruction quality criteria
as described below. Multiple signal candidates in the event are allowed.

The photon candidates are EMC clusters in the angular
region $0.41 < \theta  < 2.41$ radians where $\theta$ is the polar angle
with respect to the beam axis in the laboratory frame.
The lateral energy distribution ($LAT$) \cite{LAT}
measures the transverse energy profile of a cluster;
requiring this to be less than 0.5 suppresses clusters
due to both electronic noise and hadronic interactions.
The azimuthal asymmetry of the energy deposition in a cluster is
measured by the $A_{42}$ Zernike moment \cite{A42}.
Requiring $A_{42}$ less than 0.1 further rejects clusters
from hadronic interactions. In addition,
the angular separation between the direction of the candidate
and of any charged track in the event should be at least 9$^\circ$ 
in the laboratory frame (split-off rejection).
The clusters satisfying these criteria come mostly from $\pi^0$ decay.
We reject photon candidates that, when
combined with any other photon, produce a mass between
114 MeV and 146 MeV ($\pi^0$ veto). The partner photon
must have energy greater than 30 MeV and $LAT$ $<$ 
0.8 without any requirement on $A_{42}$ and split-off rejection.

Backgrounds arise from combinatorial background in $B$ decays and
continuum events, and 
decays of $\psi(2S)$ mesons produced either promptly or in ISR events.
To suppress $B$-background contributions,
we require
$p^*_{\chi_c} > 2.0$ GeV
and
$p^*_{J/\psi} > 2.0$ GeV.
For the $\chi_c$ control sample,
we require
$p^*_{\chi_c} < 1.7$ GeV
and
$p^*_{J/\psi} < 2.0$ GeV.
The combinatorial background 
for $J/\psi$ candidates in continuum events
is reduced by requiring $|\cos\theta_H^{J/\psi}| < 0.9$ \cite{babardbl}, where
$\theta_H^{J/\psi}$ is the $J/\psi$ helicity angle, measured in the rest frame of the
$J/\psi$, between the positively charged lepton daughter and the $\gamma J/\psi$ system.

The backgrounds from prompt $\psi(2S)$ radiative decay to $\gamma\chi_c$ are
indistinguishable from the signal. The estimated contribution from prompt $\psi(2S)$ production
will be subtracted from the measured cross-sections.

Substantial backgrounds are due to ISR production of $\psi(2S)$
decaying to $\gamma \chi_c$, 
which produces low multiplicity and a jet-like event shape.
To suppress such backgrounds,
the ratio $R_2$ of second and zeroth Fox-Wolfram moments 
of the event \cite{R2} is required to be less than 0.8 and
the number of charged particles in
the event is required to be at least five ($N_{ch}$ cut).
We estimate the possible contributions from ISR production using MC samples
and subtract them from the signal yield.
Two-photon background contributions are estimated to be negligible with
all selection criteria applied.

The helicity angle of the $\gamma J/\psi$ system ($\theta_H$) 
is the angle, measured in the rest frame of the $\gamma J/\psi$ system,
between the momentum of the $J/\psi$ and the momentum of the
\ee\ center-of-mass in the laboratory frame.
The $J/\psi$ mesons from combinatorial background tend to be along
the direction of the boost vector which makes
$\cos\theta_H$ close to unity whereas the distribution of signal events is
flat.
We optimize the $\cos\theta_H$ cut using MC samples
by maximizing the figure of merit 
$N_{sig}^2/(N_{cont}+N_{B\overline B})$, where $N_{sig}$,
$N_{cont}$ and $N_{B\overline B}$ are the numbers of events
from signal, continuum, and $B\overline B$ background expected in the data sample respectively.
The scale of $N_{sig}$ is not sensitive to the optimized cut.
For $N_{cont}$ we use the yield from off-resonance data
multiplied by $({\cal L}_{on}+{\cal L}_{off})/{\cal L}_{off}$.
The optimized cut is found to be $\cos\theta_H$ $<$ 0.4. The same cut is applied for
the $X(3872)$ search which has similar kinematics.

We extract the signal yield using an unbinned maximum likelihood (UML) fit (nominal fit)
for the distribution of $\Delta M$, the mass difference between the signal $\chi_c$ or $X(3872)$
candidate and the daughter $J/\psi$ candidate.
We use a $\Delta M$ range
[0.25, 0.60] GeV for the $\chi_{c}$ searches and
[0.60, 0.95] GeV for the $X(3872)$ search in the nominal fit.
To estimate the systematic uncertainty, we use
[0.25, 0.35] GeV and [0.50, 0.60] GeV as sideband regions and
[0.35, 0.50] GeV as the core signal region for the $\chi_c$ states.

The $\Delta M$ distribution for signal candidates is described 
by a Crystal Ball Line shape (CBL)
which is a Gaussian (described by the peak value $\Delta M_0$
and resolution $\sigma_{\Delta M}$) with a power law tail
$1/(\Delta M_0 - \Delta M + {\rm const})^n$, at a value of
$\Delta M_0 - \alpha\cdot\sigma_{\Delta M}$.
We use different PDFs
for $\chi_{c1}$, $\chi_{c2}$, and $X(3872)$,
averaged over the $e^+e^-$ and $\mu^+\mu^-$ modes.
The parameter values used in the CBL are determined using MC simulation
and are then fixed in the nominal fit.
The resolution $\sigma_{\Delta M}$ is 14.0 MeV, 15.3 MeV and 20.5 MeV for
$\chi_{c1}$, $\chi_{c2}$, and $X(3872)$ respectively and these are scaled by $\beta$,
a scale factor for the $\Delta M$ resolution. 
The mean $\Delta M_0$ for each of $\chi_{c1}$, $\chi_{c2}$, and $X(3872)$
is given by the known mass shifted by $\delta$, an offset of
the PDF in $\Delta M$. 
The difference of the $\chi_{c1}$ and $\chi_{c2}$ masses is constrained to
the known value 45.5 MeV \cite{PDG}.
The $\beta$ and $\delta$
parameters are determined as
$\beta = 0.89 \pm 0.03$ and $\delta = (2.7 \pm 0.4)$ MeV 
using a control data sample of $\chi_c$ mesons from $B$ decay
and fixed in the nominal fit.
The background line shape is described
by a third-order Chebyshev polynomial with free coefficients.

\begin{figure}[t]
\centering
\includegraphics[width=5.3cm]{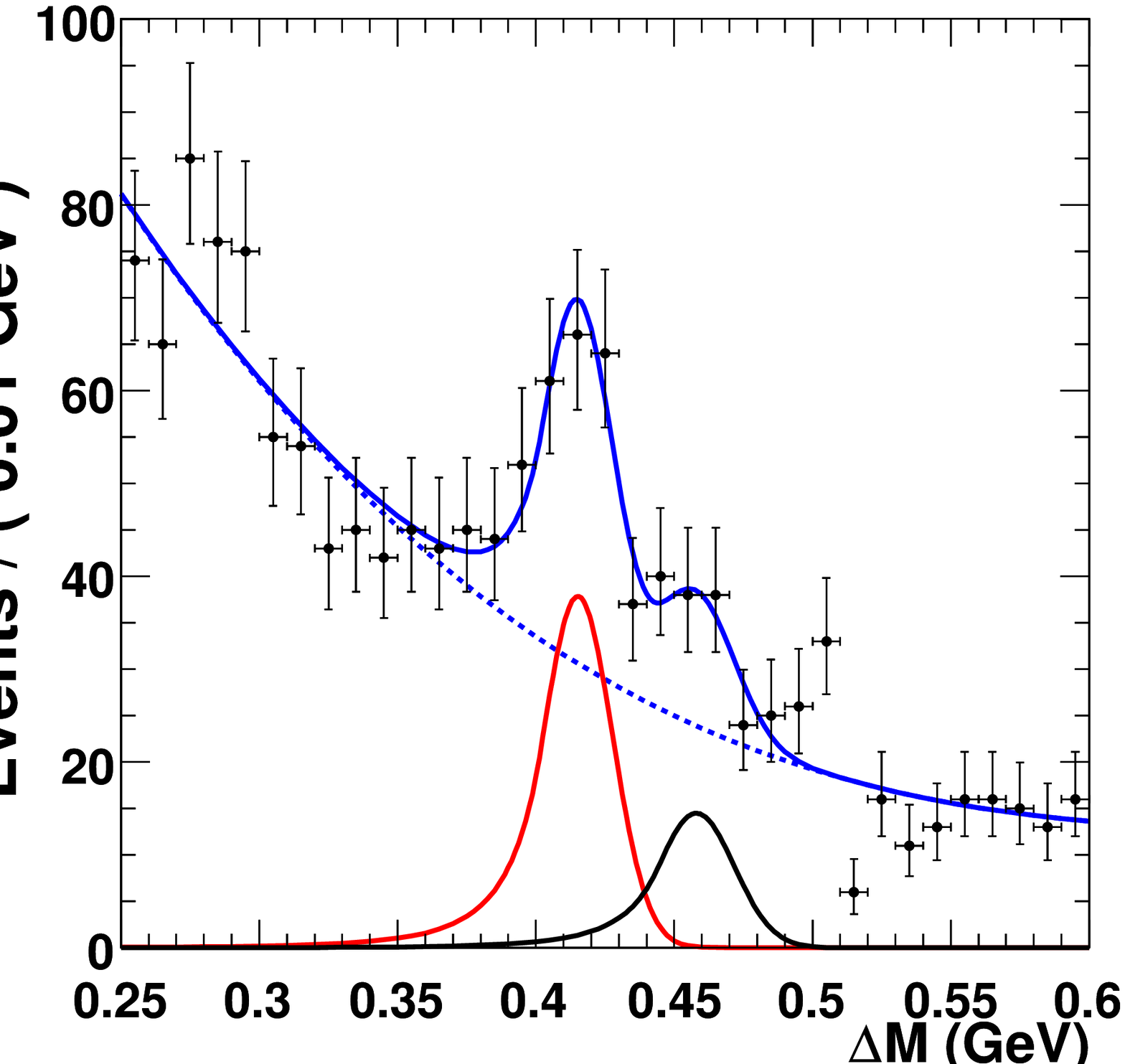}
\includegraphics[width=5.3cm]{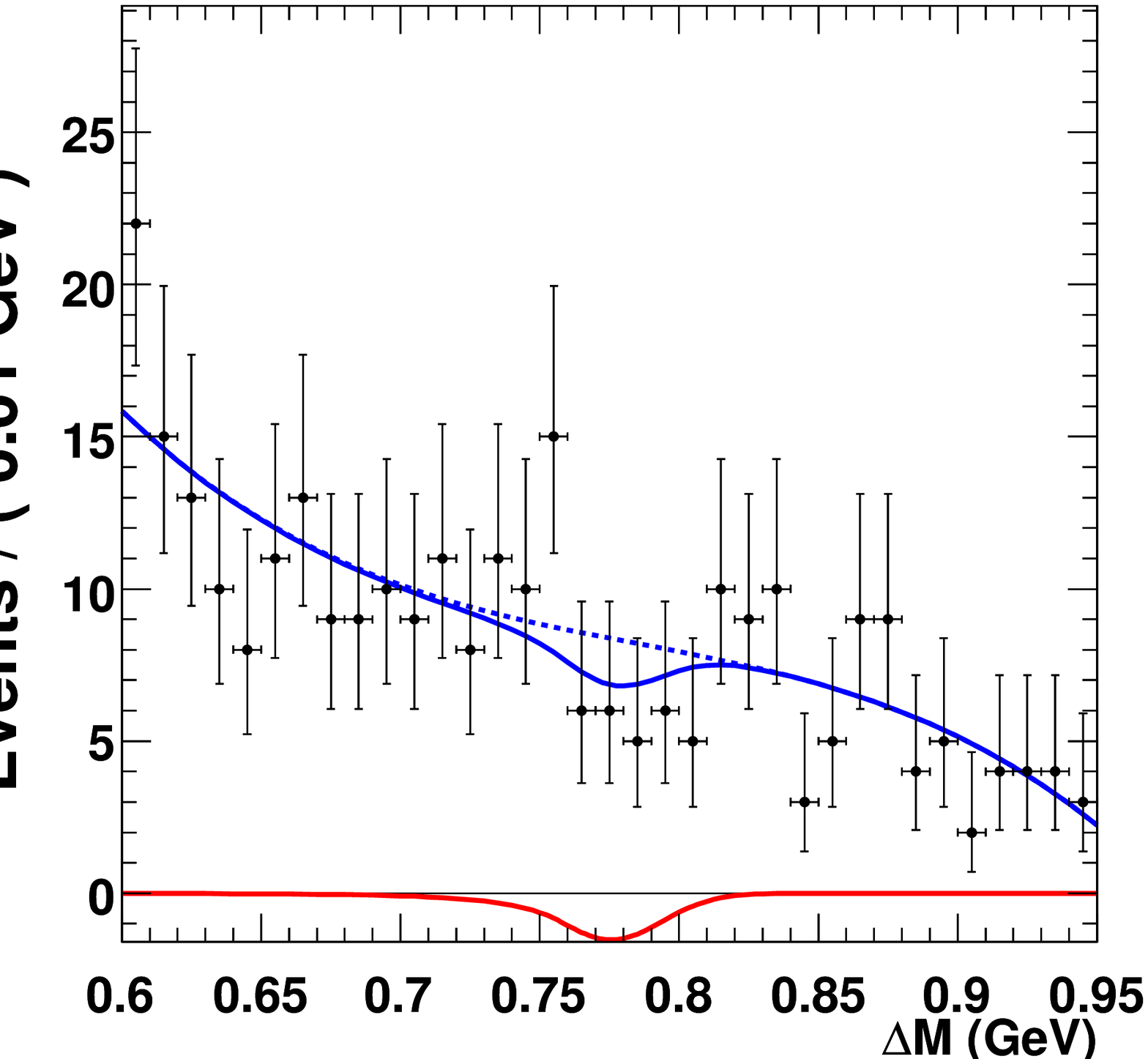}
\caption{(color online) Nominal fit result for the $\chi_c$ search (top) and
  the $X(3872)$ search (bottom) from the 386 fb$^{-1}$ data sample ($p^*>2.0$ GeV).
  The points represent the data, the dashed lines are background PDFs, the
  solid lines below the points are signal PDFs, and
  the solid lines on the points are the total PDFs.
  } 
\label{fig:fitresult}
\end{figure}

The results for the nominal fit are presented in Fig. \ref{fig:fitresult}.
For the $\chi_{c1}$ and $\chi_{c2}$ searches, we analyze 1417 events after all selection criteria.
The number of $\chi_{c1}$ candidates is
$134 \pm 23$ and 
the number of $\chi_{c2}$ candidates is
$56 \pm 19$.
For the $X(3872)$ search, we find $N_{X(3872)} = -8 \pm 11$ 
from 293 events.

The ISR $\psi(2S)$ backgrounds are estimated using MC samples to be
9.4 events for $\chi_{c1}$ and 5.1 events for $\chi_{c2}$.
Subtracting these from the fitted yields we find
$N_{\chi_{c1}} = 125 \pm 23$ and $N_{\chi_{c2}} = 51 \pm 19$,
which we attribute to the sum of prompt $\chi_c$ production and feed-down
from prompt $\psi(2S)$ production.

To estimate the signal detection efficiency $\epsilon$,
we decompose it into three factors: efficiencies of reconstruction
($\epsilon_r$),
$\pi^0$ veto (\vetoEff) and split-off rejection (\splitoffEff).
The efficiency becomes smaller in low $p^*$ bins and high $\cos\theta^*$ bins
owing to the $p^*_{J/\psi} > 2.0$ GeV requirement and lower detector coverage
near the endcap region.
To get an estimate of $\epsilon_r$, we divide the region $2.0 < p^* < 5.0$ GeV into 6 bins and
$-1.0 < \cos\theta^* < 1.0$ into 5 bins.
We correct using the formula
$\epsilon_r = w^{p^*}_{i}\ \epsilon_{ij}\ w^{\cos\theta^*}_{j}$
where $w^{p^*}_{i}$ and $w^{\cos\theta^*}_{j}$ are weights and
$\epsilon_{ij}$ is an efficiency matrix ($i=$1,6; $j=$1,5), averaged over the $e^+e^-$
and $\mu^+\mu^-$ modes using single particle MC samples.
The weights are defined by
$$w_i = \frac{N^{p^*}_i/\epsilon^{p^*}_i}{\displaystyle \sum_{k=1}^6 N^{p^*}_k/\epsilon^{p^*}_k}, \ \ \ w_j = \frac{N^{\cos\theta^*}_j/\epsilon^{\cos\theta^*}_j}{\displaystyle \sum_{k=1}^5 N^{\cos\theta^*}_k/\epsilon^{\cos\theta^*}_k},$$
where $\epsilon^{p^*}_i$ and $\epsilon^{\cos\theta^*}_j$ are efficiencies in bins of $p^*$ and $\cos\theta^*$, determined from the single $\chi_c$ MC samples, and
$N^{p^*}_i$ and $N^{\cos\theta^*}_j$ are the yields in each bin, extracted from the binned fit to the data sample.
For the $X(3872)$ search, we use the averaged efficiency when the weights for
$\chi_{c1}$ and $\chi_{c2}$ are used, because of the limited statistics
for the number of $X(3872)$ candidates.
The $\epsilon_{ij}$ values are determined from the single $X(3872)$ MC sample.
With these corrections, the $\epsilon_r$ values are 10.1\%, 9.3\%, and 8.4\%
for $\chi_{c1}$, $\chi_{c2}$, and $X(3872)$, respectively.

To estimate \vetoEff\ and \splitoffEff,
we need to have knowledge of the efficiency
as a function of photon ($N_\gamma$) or charged track multiplicity ($N_{ch}$), and
the $N_\gamma$ or $N_{ch}$ fractional distribution of signal events,
because \vetoEff\ and \splitoffEff\ are strongly dependent on the number of photons or charged tracks
in the event.
We estimate efficiencies for each $N_\gamma$ and $N_{ch}$ bin using signal MC simulation
corrected by the
data-to-MC difference using $\chi_c$ candidates from $B$ decays.
The distributions of $N_\gamma$ and $N_{ch}$ for signal events are estimated
from the sideband-subtracted data sample.
The $N_\gamma$ distribution ranges from 1 to 18 and the $N_{ch}$ distribution ranges from 5 to 14.
We estimate \vetoEff$=0.80$ and \splitoffEff$=0.96$
from an average calculated by the following formula:
$$\epsilon = \frac{\displaystyle \sum_i N_{pi} \cdot \epsilon(N_{i}) }{\displaystyle \sum_j N_{pj}}
= \frac{\displaystyle \sum_i N_{oi}}{\displaystyle \sum_j N_{pj}}$$
where $N_{pi}$ stands for the number of photons or charged tracks produced
in the $i$th bin, $N_{oi}$ for the number of photons or charged tracks observed
in the $i$th bin, $N_{pi} = N_{oi} / \epsilon(N_{i})$, and
$\epsilon(N_{i})$ is the efficiency of the $i$th bin in the distribution of $N_{\gamma}$ or $N_{ch}$.
For the $X(3872)$ search, we use the same \vetoEff\ and \splitoffEff\ as for the $\chi_c$.
The total efficiency $\epsilon$ is the product $\epsilon_r \cdot$ \vetoEff
$\cdot$ \splitoffEff\ and is
estimated to be 7.7\%, 7.1\% and 6.4\%, respectively, for the $\chi_{c1}$, $\chi_{c2}$, and $X(3872)$.

\begin{table}[h]
\centering
\caption{Systematic uncertainties (quoted in \%) on $\sigma_{N_{ch}>2}$ defined in the text. 
  \label{tab:syst}}
\begin{tabular}{c|ccc} 
           &\ \ $\chi_{c1}$ &\ \ $\chi_{c2}$ &\  $X(3872)$\\\hline
$p^*$/$\cos\theta^*$ correction  &\ \ \  13.3 &\ \ 26.5 &  34.9 \\
Track efficiency                 &\ \ \   0.5 &\ \  0.5 & 0.5\\
Charged PID                      &\ \ \   7.2 &\ \  7.2 & 7.2\\
Photon PID                       &\ \ \   1.8 &\ \  1.8 & 1.8\\
$\pi^0$ veto efficiency ($\epsilon_v$)         &\ \ \   2.3 &\ \  2.3 & 2.3\\
Split-Off rejection efficiency ($\epsilon_s)$   &\ \ \   0.4 &\ \  0.4 & 0.4\\
PDF                              &\ \ \   3.5 &\ \ 11.2 & 15.1 \\
ISR Background                   &\ \ \   3.8 &\ \  5.0 & --   \\
$\prod {\cal B}_i$               &\ \ \   5.4 &\ \  5.0 & 0.7  \\\hline
Total                            &\ \ \  17.1 &\ \ 30.6 & 38.8 \\
\end{tabular}
\end{table}

The sources of systematic uncertainty are summarized in Table~\ref{tab:syst}.  
The dominant uncertainty is from the reconstruction efficiency ($\epsilon_r$)
correction from the $p^*$ and $\cos\theta^*$ distributions.
For the $\chi_c$ search, we assign the systematic uncertainty as
the r.m.s. spread of 10,000 simulated experiments
(each experiment gives one $\epsilon_r$ value)
with weights generated according to the central values and errors from
the $p^*$ and $\cos\theta^*$ binned fit results.
For the $X(3872)$ search, we adopt a conservative approach. We calculate
separately the r.m.s. values corresponding to the binned fit results
for $\chi_{c1}$ and $\chi_{c2}$, and assign the sum of r.m.s. values
as the systematic uncertainty for the $X(3872)$ reconstruction efficiency.

The error from the PDF modeling is estimated by a quadratic sum over the changes in the yield
from an alternative background line shape $e^{-[p_0 + p_1 (\Delta M) + p_2 (\Delta M)^2]}$,
and $\pm 1$ standard deviation of the uncertainties in the measured $\delta$ and $\beta$
in the $\chi_c$ control sample from $B$ decay.
We take the data-to-MC difference in track reconstruction efficiency as a source of
systematic uncertainty.
To estimate systematic uncertainties in charged PID efficiencies, we assign
the difference when taking $\pm 1$ standard deviation
of each error depending on momentum and azimuthal angle of tracks measured using control samples.
The systematic uncertainty of photon identification is estimated by comparing data
with MC simulations of $\tau^+ \rightarrow \pi^+\nu$ and $\tau^+ \rightarrow \rho(\pi^+\pi^0)\nu$ samples.
We assign half of the ISR $\psi(2S)$ background estimate as systematic
uncertainty for the $\chi_c$ search. The uncertainty of the ISR background
is neglected for the $X(3872)$ search.
The $\prod {\cal B}_i$ is a product of sub-decay mode branching fractions, that is
${\cal B}(\chi_c\rightarrow \gamma J/\psi) \cdot
[{\cal B}(J/\psi \rightarrow e^+e^-) + {\cal B}(J/\psi \rightarrow \mu^+\mu^-)]$.
The systematic error related to $\prod {\cal B}_i$ is estimated from the reference values~\cite{PDG}.
The systematic uncertainties from $\epsilon_v$ and $\epsilon_s$ evaluations are estimated by
a quadratic sum over the deviations in two cases: when the data-to-MC correction is not used
and when $N_\gamma$ and $N_{ch}$ distributions are taken without sideband subtraction to see the effect of backgrounds on the distribution.

Table~\ref{tab:R18finalresult} summarizes the measurements and all the quantities we need to calculate
$\sigma_{N_{ch}>2}$, that is
the cross-section of prompt $\chi_c$ or $X(3872)$ production ($\sigma (e^+e^-\rightarrow c\bar c X)$)
times the probability of the rest of the event ($X$) having more than two charged tracks,
${\cal P}_{N_{ch} >2}$.
The result $\sigma_{N_{ch}>2}$ is derived from the formula
$N_{sg} = {\cal L} \cdot \epsilon \cdot \sigma_{N_{ch}>2} \cdot \prod {\cal B}_i$
where $N_{sg}$ is the number of $\chi_c$ or $X(3872)$ candidates from $e^+e^-$ annihilation.
In the case of $\chi_c$, $\sigma_{N_{ch}>2}$
includes the prompt $\psi(2S)$ feed-down contribution. For the $X(3872)$, we measure the product
$\sigma_{N_{ch}>2} \cdot {\cal B}(X(3872)\rightarrow \gamma J/\psi)$
because the $X(3872)\rightarrow \gamma J/\psi$ BF is unknown.
    
\begin{table}[h]
  \begin{center}
\caption{Signal yield $N_{sg}$ from the nominal fit after subtracting the ISR $\psi(2S)$ estimate;
  signal detection efficiency ($\epsilon=\epsilon_r \cdot \epsilon_{v} \cdot \epsilon_{s}$);
  product of sub-decay mode BF's ($\prod {\cal B}_i$);
  integrated on- and off-resonance luminosity ($\cal L$);
  $\sigma_{N_{ch} >2}$ (defined in the text) and 
  its upper limit including systematic uncertainties;
  $\sigma^{prompt}_{N_{ch} >2}$ ($\sigma_{N_{ch} >2}$ for the prompt production)
  and its upper limit including systematic uncertainties.
  Upper limits are at the 90\% C.L. Note that $\sigma_{N_{ch} >2}$ for $X(3872)$ denotes
  $\sigma_{N_{ch}>2} \cdot {\cal B}(X(3872)\rightarrow \gamma J/\psi)$.
}
\label{tab:R18finalresult}
\begin{tabular}{c|ccc}  
                   & $\chi_{c1}$  & $\chi_{c2}$ & $X(3872)$      \\\hline
$N_{sg}$   & 125$\pm$23   & 51$\pm$19   &--8.0$\pm 11$ \\
  &                & ($<75$)     & ($<15$)      \\
$\epsilon$ (\%)      &  7.7 &  7.1        & 6.4 \\
$\prod {\cal B}_{i}$ (\%)    &  4.2 &  2.4        & 11.9\\
$\cal L$ (fb$^{-1}$) & 386  & 386         & 386 \\\hline\hline
 $\sigma_{N_{ch}>2}$ (fb)         &\ \ 99$\pm 18 \pm 17$\ \ &\ \ 78$\pm 28 \pm 24$\ \ &--3$\pm 4 \pm 1$  \\
                        &                \ \  &\ \ $<125 $              \ \  & $<5$ \\\hline
 $\sigma^{prompt}_{N_{ch} >2}$ (fb)&\ \ 41$\pm 18 \pm 21$ &$23 \pm 28 \pm 26$     &--3$\pm 4 \pm 1$  \\
   &   $<77$   &$<79 $& $<5$ \\\hline
  \end{tabular}
  \end{center}
\end{table}

For prompt $\chi_c$ production, 
it is necessary to subtract prompt $\psi(2S)$ feed-down to $\chi_c$.
The contribution of prompt $\psi(2S)$ production is estimated to be
(58 $\pm$ 12) fb for $\chi_{c1}$ and (54 $\pm$ 11) fb for $\chi_{c2}$ using
$\sigma(e^+e^-\rightarrow \psi(2S) X) = (0.67 \pm 0.13)$ pb for $p^* > 2.0$ GeV
\cite{belleprompt} and the $\psi(2S)\rightarrow\gamma\chi_{c}$ BF \cite{PDG}.
The errors are included as systematic uncertainties in the prompt $\chi_{c}$ production cross-section.
Feed-down from other $\psi(2S)$ decay modes with photons is
checked using MC simulation:
$J/\psi \pi^0\pi^0$, $J/\psi \eta(\gamma\gamma)$, $J/\psi \eta(\gamma\pi^+\pi^-)$,
$J/\psi \eta(\pi^+\pi^-\pi^0)$, and $J/\psi \eta(\pi^0\pi^0\pi^0)$. No background from these decays is seen in the MC simulation.
The resultant cross-sections, $\sigma_{N_{ch} > 2}^{prompt}$, for $\chi_c$ production are shown in Table~\ref{tab:R18finalresult}.

Our measurements use an additional kinematic cut $p^*_{\chi_c} > 2.0$ GeV
which has little effect on the cross-section because leading-order
contributions are from two-body $e^+e^-$ annihilation processes.
To compare these results with the theoretical predictions in Ref. \cite{theory}, 
the value of ${\cal P}_{N_{ch} >2}$ should be estimated correctly.
Nevertheless, our upper limits
are comparable with the NRQCD cross-section predictions.

In summary, we have searched for prompt production of
$\chi_{c1}$ and $\chi_{c2}$ 
in \ee\ annihilation near $\sqrt{s} = 10.6$ GeV.
We observe candidates for these $\chi_c$ states, but the measured
cross-sections are compatible, within statistics, with the expected
contributions of $\chi_c$ feed-down from prompt $\psi(2S)$ production.
The 90\% confidence level upper limits on $\sigma^{prompt}_{N_{ch} >2}$ are
77 fb for $\chi_{c1}$ and 79 fb for $\chi_{c2}$
with $p^\ast_{\chi_c} > 2.0$ GeV.
We find no evidence for prompt $X(3872)$ production via the decay
$X(3872)\rightarrow \gamma J/\psi$. We set the 90\% confidence lever upper limit
on $\sigma_{N_{ch}>2} \cdot {\cal B}(X(3872)\rightarrow \gamma J/\psi)$ to be 5 fb.
The upper limits presented on prompt $\chi_c$ production are
significant improvements on the previously reported results \cite{belleprompt}.
These limits are comparable to the theoretical cross-section predictions of Ref.~\cite{theory}.
Upper limits on prompt production of $\chi_c$ in comparison with $J/\psi$ and $\psi(2S)$
prompt production \cite{babarprompt,belleprompt},
can be used to further our understanding of
the charmonium prompt production mechanism \cite{prompt_theory,dbltheory,theory}.

We are grateful for the excellent luminosity and machine conditions
provided by our \pep2\ colleagues, 
and for the substantial dedicated effort from
the computing organizations that support \babar.
The collaborating institutions wish to thank 
SLAC for its support and kind hospitality. 
This work is supported by
DOE
and NSF (USA),
NSERC (Canada),
CEA and
CNRS-IN2P3
(France),
BMBF and DFG
(Germany),
INFN (Italy),
FOM (The Netherlands),
NFR (Norway),
MIST (Russia),
MEC (Spain), and
PPARC (United Kingdom). 
Individuals have received support from the
Marie Curie EIF (European Union) and
the A.~P.~Sloan Foundation.

\renewcommand{\baselinestretch}{1}


\begin{thebibliography}{99}

\bibitem{prompt_theory} G. A. Schuler, Eur. Phys. J. C {\bf 8}, 273 (1999); F. Yuan, C.-F. Qiao, and K.-T. Chao, Phys. Rev. D {\bf 56}, 321 (1997).
\bibitem{dbltheory} K.-Y. Liu, Z.-G. He, and K.-T. Chao, hep-ph/0408141; Y. Iwasaki, Phys. Rev. D {\bf 16}, 220 (1977).
\bibitem{babarprompt} \babar\ Collaboration, B. Aubert {\it et al.}, Phys. Rev. Lett. {\bf 87}, 162002 (2001).
\bibitem{belleprompt} Belle Collaboration, K. Abe {\it et al.}, Phys. Rev. Lett. {\bf 88}, 052001 (2002).
\bibitem{belledbl} Belle Collaboration, K. Abe {\it et al.}, Phys. Rev. D {\bf 70} 071102 (2004);
  Belle Collaboration, K. Abe {\it et al.}, Phys. Rev. Lett. {\bf 89}, 142001 (2002).
\bibitem{babardbl} \babar\ Collaboration, B. Aubert {\it et al.}, Phys. Rev. D {\bf 72} 031101(R) (2005).
\bibitem{dbltheory2} K.-Y. Liu, Z.-G. He, and K.-T. Chao, Phys. Lett. B {\bf 557} 45 (2003); 
G. T. Bodwin, J. Lee, and E. Braaten, Phys. Rev. Lett. {\bf 90}, 162001 (2003).
\bibitem{theory} 
G. A. Schuler and M. $\rm V\ddot{a}nttinen$,
Phys. Rev. D {\bf 58} 017502 (1998).
\bibitem{x3872} \babar\ Collaboration, B. Aubert {\it et al.}, Phys. Rev. D {\bf 74} 071101(R) (2006).
\bibitem{ref:babar}
\babar\ Collaboration, B.\ Aubert {\em et al.},
Nucl.\ Instrum.\ Methods {\bf A479}, 1 (2002).
\bibitem{GEANT4}
  GEANT Collaboration, S. Agostinelli {\it et al.}, Nucl. Instrum. Methods
  Phys. Res., Sect. A {\bf 506}, 250 (2003).
\bibitem{LAT} A. Drescher {\it et al.}, Nucl. Instr. and Methods {\bf A 237}, 464 (1985).
\bibitem{A42} R. Sinkus and T. Voss, Nucl. Instr. and Methods {\bf A 391}, 360 (1997).
\bibitem{R2} G.C. Fox and S. Wolfram, Phys. Rev. Lett. {\bf 41}, 1581
(1978).
\bibitem{PDG} W.-M. Yao {\it et al.}, J. Phys. G {\bf 33} 1 (2006).  
\end{thebibliography}
\end{document}